\documentclass[]{interact}

\usepackage{epstopdf}
\usepackage[caption=false]{subfig}
\usepackage[natbibapa,nodoi]{apacite}
\theoremstyle{plain}

\theoremstyle{definition}

\theoremstyle{remark}

\usepackage{graphics}
\usepackage{array}
\usepackage{booktabs} 
\usepackage{graphicx}
\usepackage{multirow}
\usepackage{epstopdf}
\usepackage{tikz}
\usepackage{tabularx,ragged2e}
\usepackage{amsthm}
\usepackage{amsmath}
\usepackage{amsfonts}
\usepackage[ruled]{algorithm2e} 
\usepackage{pifont}
\newcommand{\cmark}{\text{\ding{51}}}
\newcommand{\xmark}{\text{\ding{55}}}

\newcommand{\ie}{\emph{i.e.,}\xspace}
\newcommand{\eg}{\emph{e.g.,}\xspace}

\newcommand{\tmmrev}[2]{{\color{#1} {#2}}}
\newcommand\xr[1]{\tmmrev{black}{#1}} 

\begin{document}

\articletype{ARTICLE TEMPLATE}

\title{The Crowd in MOOCs: A Study of Learning Patterns at Scale}

\author{
	\name{Xin Zhou\textsuperscript{a}\thanks{CONTACT Xin Zhou. Email: enoche.chow@gmail.com}, Aixin Sun\textsuperscript{a}, Jie Zhang\textsuperscript{a} and Donghui Lin\textsuperscript{b}}
	\affil{\textsuperscript{a}College of Computing and Data Science, Nanyang Technological University, Singapore; \textsuperscript{b}Faculty of Environmental, Life, Natural Science and Technology, Okayama University, Japan.}
}

\maketitle

\begin{abstract}
The increasing availability of learning activity data in Massive Open Online Courses (MOOCs) enables us to conduct a large-scale analysis of learners' learning behavior. 
In this paper, we analyze a dataset of 351 million learning activities from 0.8 million unique learners enrolled in over 1.6 thousand courses within two years.  
Specifically, we mine and identify the learning patterns of the crowd from both temporal and course enrollment perspectives leveraging mutual information theory and sequential pattern mining methods. 
From the temporal perspective, we find that the time intervals between consecutive learning activities of learners exhibit a mix of power-law and periodic cosine function distribution.
By qualifying the relationship between course pairs, we observe that the most frequently co-enrolled courses usually fall in the same category or the same university. 
We demonstrate these findings can facilitate manifold applications including recommendation tasks on courses. A simple recommendation model utilizing the course enrollment patterns is competitive to the baselines with 200$\times$ faster training time.
\end{abstract}

\begin{keywords}
MOOC; learning patterns; mining sequences; recommender systems
\end{keywords}

\section{Introduction}
Following the booming of Massive Open Online Courses (MOOC) in 2012, many MOOC platforms have been launched in different countries.
This momentum is accelerated during the COVID-19 pandemic as classes in brick-and-mortar colleges are forced to move online~\citep{impey2021moocs}. 
Major MOOC platforms accumulate a large volume of data on learning activities, \eg video watching histories, posts in course forums, and assignment performance~\citep{qu2015visual}. 
Together with the learner profiles (\eg  genders, ages, and countries), we are offered a unique opportunity to understand the learning patterns of the crowd, at scale.

Researchers have studied the learning behavior of the crowd from various perspectives~\citep{mellati2020mooc, moore2022learner, bai2023impact}.
Specifically, authors in~\citep{hew2020predicts, moore2022learner, zhu2023mooc} evaluate learners' emotion or satisfaction in learning experiences leveraging sentiment analysis. 
Other studies~\citep{wang2015investigating, zhang2019exploring, feng2019understanding, jin2023mooc} aim to predict learners' dropout or completion of courses based on their learning activities.
The research community is also interested in investigating the learners' demographics on MOOCs~\citep{bayeck2016exploratory, aljohani2021learners}.
A recent work~\citep{ruiperez2022large} performs a large-scale analysis on the regional patterns across 15 different MOOC platforms.
However, most of the existing studies either depend on only a few available MOOC courses, or they only focus on a shallow analysis of the learners from their demographics (\eg country of origin, gender) or preferences.
\xr{
Previous studies investigating learning patterns within MOOCs often suffer from limitations in generalizability due to the use of restricted course sets~\citep{boroujeni2019discovery, kokocc2021unfolding}. Additionally, existing approaches that leverage learning logs may not fully capture the underlying learning patterns. This paper addresses this gap by leveraging a large-scale MOOC dataset encompassing hundreds of millions of learning activities. Through quantitative analysis, we aim to uncover the learning patterns of learners at scale.
}
\xr{Specifically, we employ a multifaceted analytical approach to uncover learning patterns. First, we leverage statistical analysis to extract temporal trends in learner engagement on the MOOC platform. This analysis sheds light on when learners are most active. Secondly, we delve deeper into the dataset using mutual information theory and sequential pattern mining techniques.  These techniques elucidate the co-enrollment patterns exhibited by learners, providing insights into how learners select and enroll in courses together.
}

\xr{We state that the findings presented in this study hold significant implications for various stakeholders within the MOOC ecosystem, including platform operators, course instructors, and learners themselves. As a case study detailed in Section~\ref{sec:recom}, we demonstrate the utility of sequential enrollment patterns for learner course recommendations. Our proposed recommendation model, which leverages these co-enrollment patterns, achieves competitive performance compared to baseline methods while exhibiting a substantial reduction in training time.
Furthermore, this research provides a valuable baseline for future studies investigating the impact of the COVID-19 pandemic on student learning behaviors within MOOCs, as the data employed in this study predates the global crisis.}

\section{Dataset and Descriptive Analysis}
The dataset used in this study was initially released in~\citep{feng2019understanding} for dropout prediction in MOOCs.
Specifically, this dataset contains 351,451,731 learning activities collected from 1,629 courses with 772,880 learners, ranging from August 1st, 2015 to July 31st, 2017. The courses are provided with the following information:
\begin{center}
	$\langle$course title, start date, end date, category$\rangle$.
\end{center}
It is worth noting that category information is optional and not all courses are associated with categories. 
Learners can enroll in multiple courses and the learning activities are recorded in the following format:
\begin{center}
	$\langle$course title, learner ID, activity type, timestamp$\rangle$.
\end{center}
The activity types include video (\ie watch, stop, and jump), web page (\ie check and close courseware), assignment completion (\ie submit and reset answers), and forum discussion (\ie ask or reply questions).
\begin{figure}
	\centering
	\includegraphics[trim=5 0 10 0, clip, width=0.96\linewidth]{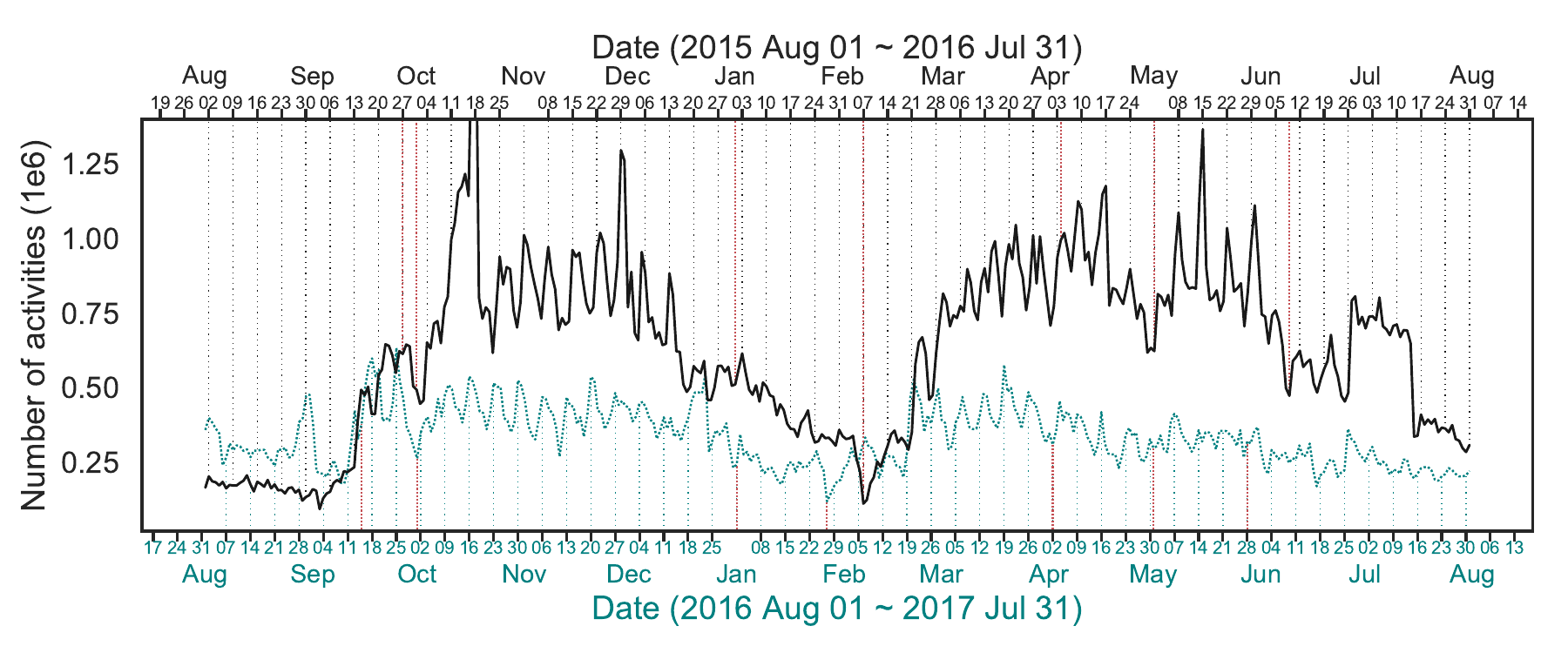}
    \vspace{-10pt}
	\caption{\xr{The distribution of learners' daily amount of activities on the platform across a two-year period. The black curve represents the first year, while the cyan curve corresponds to the second year. The dotted line highlights Sundays, and the red dashed line indicates public holidays.}}
	\label{fig:logs_daily}%
    \vspace{-10pt}
\end{figure}

In our following analysis, unless specified otherwise, we analyze the entire dataset, covering the two-year period. 
Fig.~\ref{fig:logs_daily} shows the number of daily activities in a general scene.
We observe the learners are more active during semester time than during semester breaks.\footnote{Chinese winter/summer semester breaks: 2015 (Jul. 12-Aug. 31), 2016 (Jan. 23-Feb. 22 / Jul. 13-Aug. 31), 2017 (Jan. 21-Feb. 19 / Jul. 13-Aug. 31).}
On major public holidays (\eg National Day, Lunar New Year) and best shopping events (\eg Double 11 promotion, Christmas promotion),
we find a clear decline in learning activities.
The finding is in line with our intuition that learners' behavior can be easily influenced by external events.
In our detailed week-by-week analysis of the data plots, we discern a distinct pattern in learner engagement with MOOCs. It appears that during the academic semester, learners predominantly engage with MOOCs on weekends. This trend is in stark contrast to their behavior during semester breaks, where we observe heightened activity on weekdays. This shift in engagement patterns underscores the flexibility and adaptability of MOOC learners in balancing their learning commitments with their academic schedules.

\subsection{Research Questions}

In the prior descriptive analysis, we get a preliminary impression on the learning patterns of the crowd in MOOCs.
In this paper, we go beyond the few studies that address learners' behavior in MOOCs~\citep{matcha2019analytics, wang2019using, feng2019understanding} and perform an in-depth view of the learning pattern to answer the following research questions:
\begin{itemize}
	\item \textbf{RQ1}: Is there any temporal pattern in learning activities?
	\item \textbf{RQ2}: What are the enrollment patterns of the crowd?
	\item \textbf{RQ3}: Can the observed patterns be used for course recommendation?
\end{itemize}

To answer \textbf{RQ1}, we perform statistical analysis from multiple perspectives
to inspect whether any temporal pattern exists in learning activities.
Subsequently, we further employ correlation analysis to study the enrollment patterns of the crowd, to answer \textbf{RQ2}.
Because the course enrollments of a learner are in a temporal order, we further perform a sequential pattern mining to uncover the course enrollment patterns of the crowd.
To answer \textbf{RQ3}, we study how the observed learning patterns can benefit the task of course recommendation.
To this end, we propose one extremely simple models leveraging the findings in \textbf{RQ2}. Experiments show that our approach generates competitive performance with the baseline models.

\section{Temporal Patterns (RQ1)}
\label{sec:temporal}
\subsection{Daily/Weekly Learning Activity Patterns}
\begin{figure}
	\centering
	\subfloat[\xr{Daily pattern}\label{sfig:tmbin_daily}]{\includegraphics[trim=20 15 50 36, clip, width=0.48\textwidth]{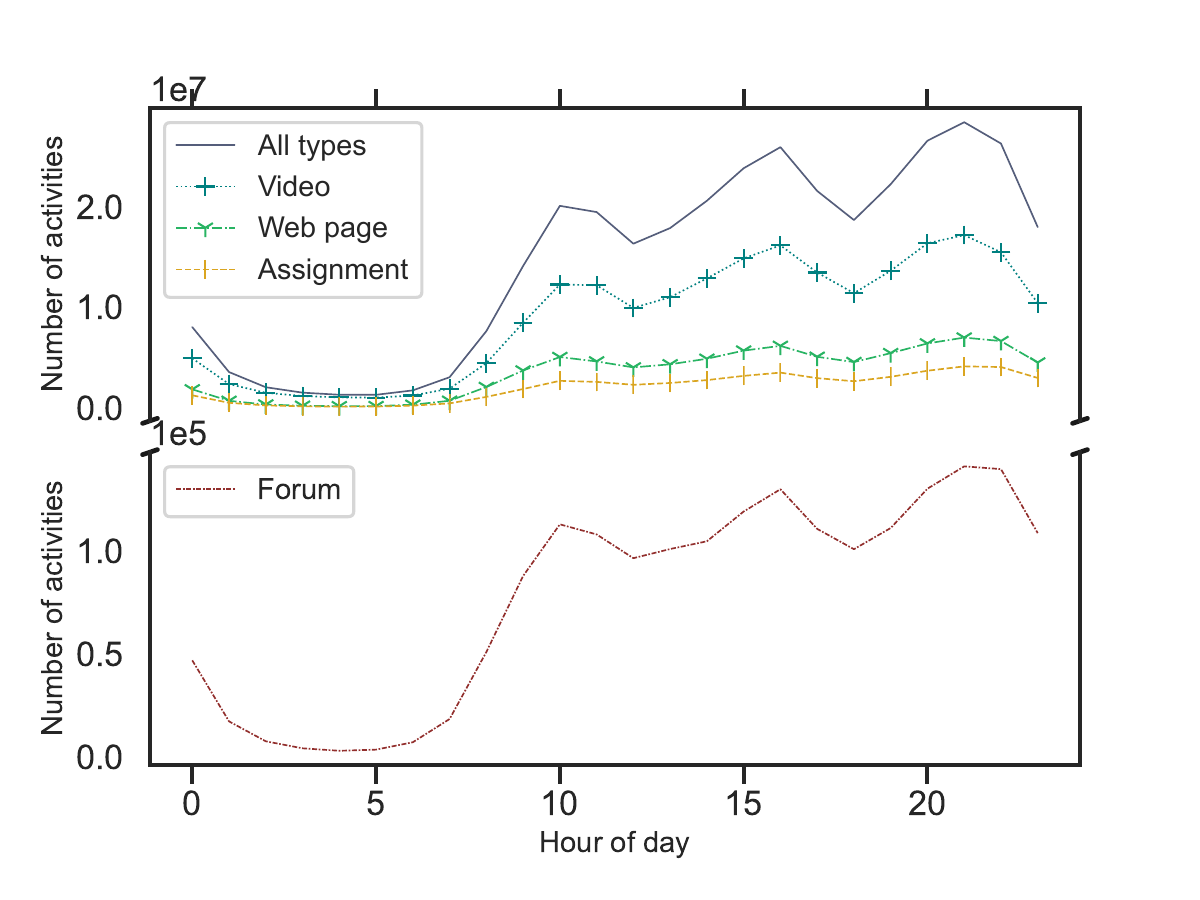}}
	\subfloat[Weekly pattern\label{sfig:tmbin_weekly}]{\includegraphics[trim=20 15 50 36, clip, width=0.48\textwidth]{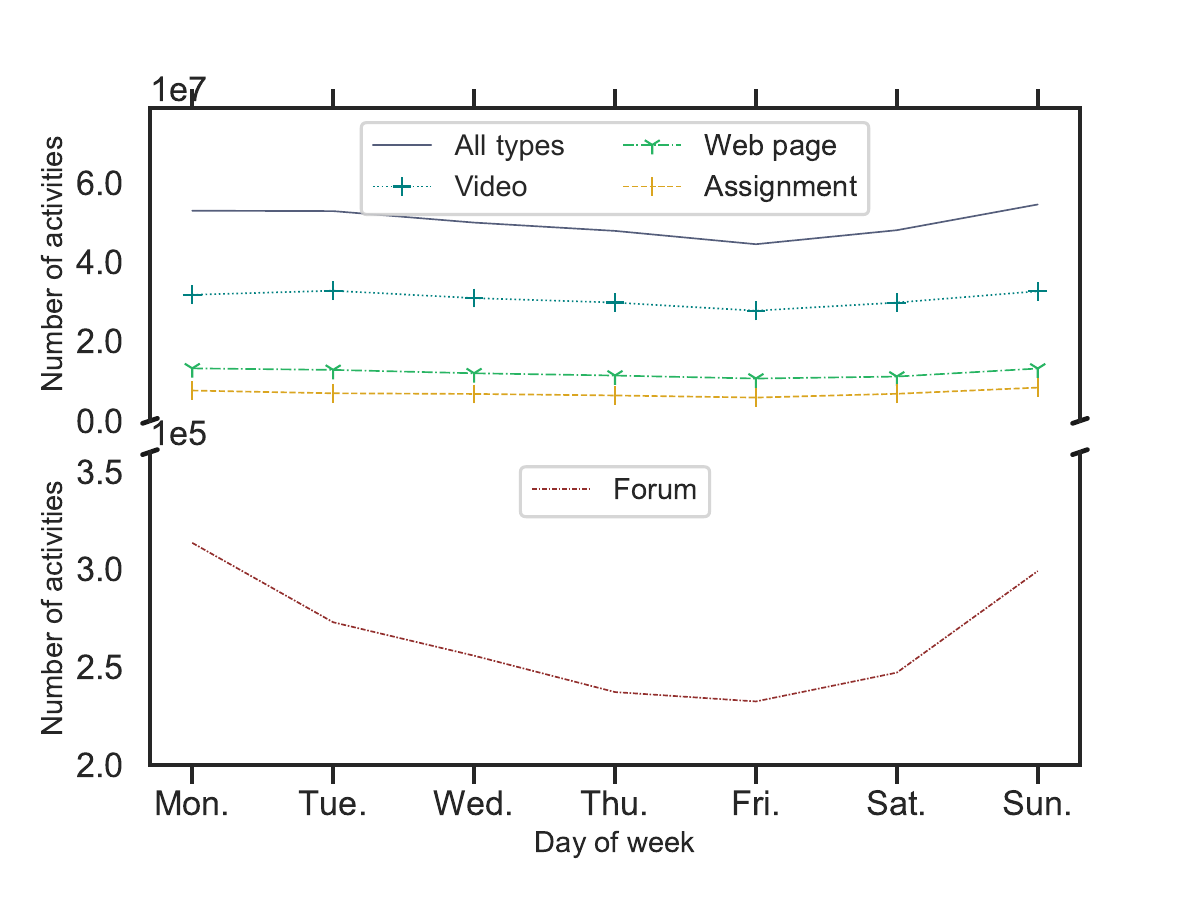}}
	\caption{\xr{Distribution of learners' activities on a daily and weekly basis.}}
   \vspace{-10pt}
	\label{fig:tmbin_lines}
\end{figure}

We explore daily and weekly patterns of learning activities from both timeline and time interval perspectives. On a daily basis, we plot the \xr{daily} pattern of learning activities in Fig.~\ref{sfig:tmbin_daily}. The \xr{daily} pattern shows an increasing trend of learning activities in all types of activities from 5:00 to 10:00. Then it follows a slight decrease before lunchtime.
The increment resumes after 12:00 and the momentum lasts until 16:00 before getting off work.
At dinner time (17:00 -- 18:00), all activities exhibit a slight decrement. However, all activities reach their daily peaks after dinner at 20:00 -- 21:00. From 21:00 to 5:00 in midnight, the activities show an opposed tendency to that of the morning.
We mark that the observed phenomenon is in line with human regularity. 

\begin{figure}[htp]
	\centering
        \vspace{-10pt}
	\includegraphics[trim=5 0 10 26, clip, width=0.7\linewidth]{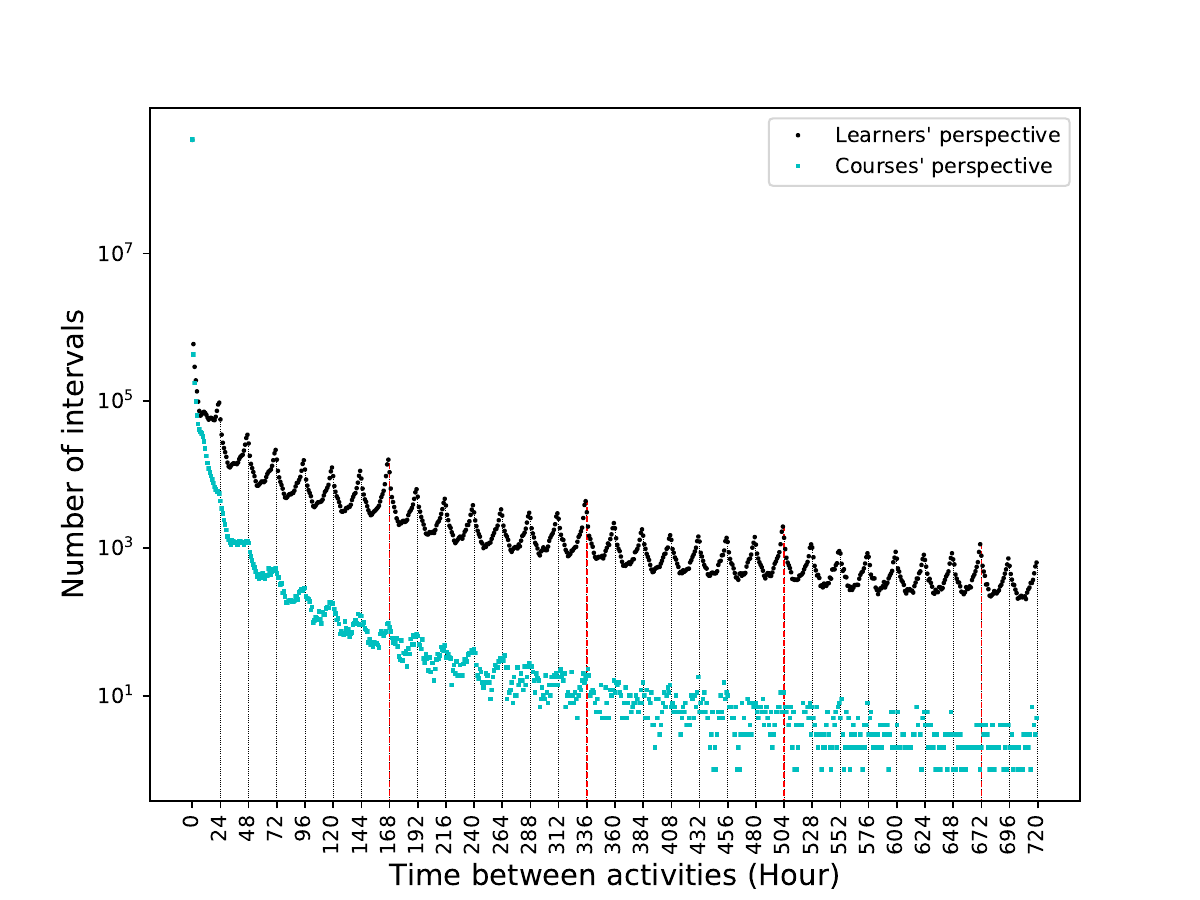}
    \vspace{-10pt}
	\caption{Distribution of the number of time intervals between consecutive activities.
		Gray dotted lines are daily indicators, and red dashed lines are weekly indicators.  $y$-axis is in log scale.}
	\label{fig:tm_interval}%
\end{figure}

Analogous to the \xr{daily} pattern, we analyze the weekly pattern in learning activities.
Different from the \xr{daily} pattern, we cannot see a wide variation in different days of a week in Fig.~\ref{sfig:tmbin_weekly}.
To sum up, the number of activities shows a downward trend in weekday and an upward trend on the weekend.
It is also worth noting that the number of different types of activities varies widely, with one or two orders of magnitude of difference.
\subsection{Consecutive Learning Activity Patterns}
Concerning the time interval between consecutive activities of learners/courses, we plot the distribution of the time intervals
between any two consecutive activities of the same learner/courses in Fig.~\ref{fig:tm_interval}.
\xr{In a learner's perspective, we define the learner engagement interval as the time elapsed between a learner's consecutive logged activities on the platform. In a course's perspective, we define the course engagement interval as the time elapsed between learners' consecutive learning activities within a specific course.}
We observe that 98.7\% of consecutive activities are conducted within an hour from
the learners' perspective. This value is as high as 99.7\% through the courses' lens.
When we dive into the distribution with time intervals limited within an hour, it shows a power law, \ie 97.5\% of consecutive activities are within 10 minutes from the learners' viewpoint.
A typical interpretation of this learning pattern is that a sequence of activities should be performed before reaching
the targeted learning material.

\xr{
As shown in Fig.~\ref{fig:tm_interval}, when the peaks are aligned, both distributions exhibit a power-law behavior, consistent with observations in many other social networks~\citep{barabasi1999emergence}.}
However, we find some exceptions in the consecutive activities of learners.
Besides the daily pattern of the activities, 
we see a slight increase in the number of consecutive activities when the time interval is exactly an integer multiple of the week.
That is, learners' consecutive activities
also show a weekly pattern in time intervals. We capture the distribution with a mix of power-law and a periodic cosine function:
\begin{equation}\label{equ:tmi}
	f(t) = k t^{-\alpha} - \left \lvert A\times cos \left (\frac{1}{T} \pi t - \frac{\pi}{2} \right ) \right \rvert,
\end{equation}
where $k, A$ are constant parameters called the scaling factor and amplitude, respectively. $\alpha$ is the law's exponent.
The latter half of the above function captures the \xr{daily} pattern, indicating both distributions oscillate with a period of $T=24$ hours along with the power law.

\xr{
The \xr{daily} and weekly patterns identified in 
Research Question 1 (RQ1)  support the notion of temporal stability and consistency in human behavior~\citep{diener2009temporal, funder1991explorations}. This aligns with established psychological principles, which posit that individuals tend to exhibit consistent behavioral patterns under recurring conditions~\citep{ouellette1998habit}. In the context of MOOCs, this translates to learners being more likely to engage in learning activities at consistent times.
}
Fig.~\ref{fig:tm_interval} also reveals that the number of intervals related to courses shows faster decay than that of learners. 
This observation indicates that less active courses are more likely to be ignored by learners, along the time.

\section{Course Enrollment Patterns (RQ2)}
\label{sec:enrollment}
In this section, we address our second research question: \textit{What are the enrollment patterns of the crowd?}
We start by employing the Jaccard similarity index and pointwise mutual information to measure how courses are likely co-enrolled.
We further investigate the course co-enrollment and transition patterns at the category level to reveal the relationship between categories.
Finally, inspired by the temporal enrollment activities of learners, we organize the course enrollment of each learner as a sequence and perform sequential pattern mining to discover frequent patterns between courses.

\begin{figure}
	\centering
	\subfloat[Jaccard similarity index ]{\includegraphics[trim=10 28 10 13, clip, width=0.43\linewidth]{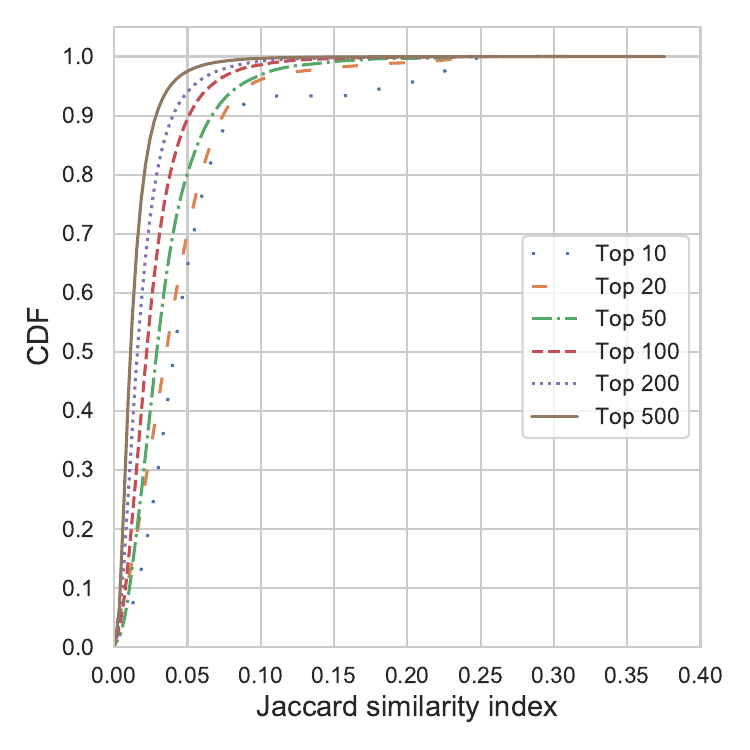}}
	\subfloat[Network topology ]{\includegraphics[trim=5 130 10 140, clip, width=0.42\linewidth]{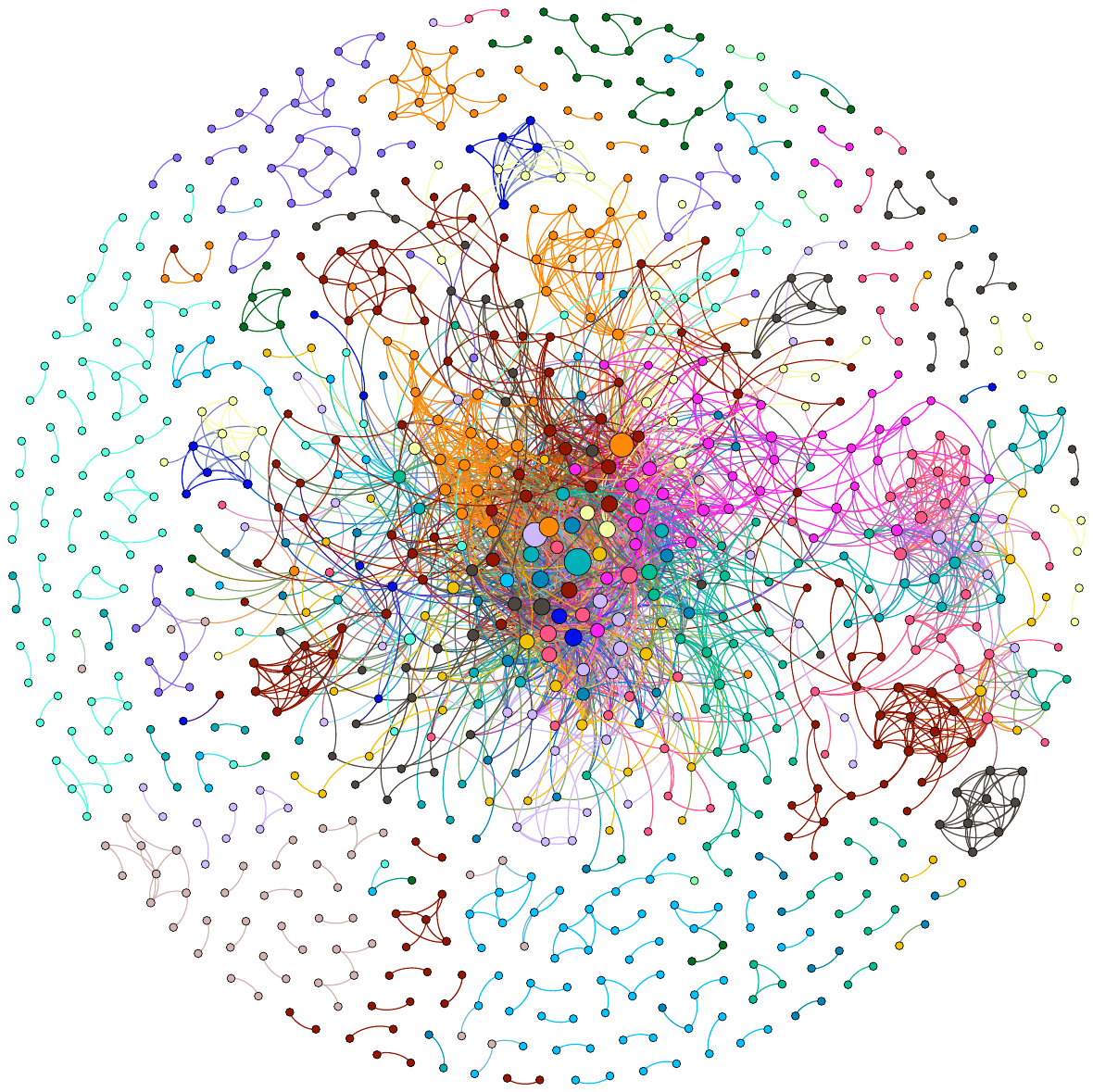}}
   \vspace{-10pt}
   \caption{Jaccard similarity index and the network topology of co-enrolled courses. In the network topology plot, each node denotes a course. Node size is proportional to the number of enrollment of the course, and nodes with the same color are within the same course category.}
   \vspace{-10pt}
	\label{fig:coenroll_cdfV4}%
\end{figure}

Let the number of learners enrolled to both courses $c_i$ and $c_j$ be
$\mathbb{L}(c_i \cap c_j)$, and the number of learners enrolled to either 
$c_i$ or $c_j$ be $\mathbb{L}(c_i \cup c_j)$. We compute Jaccard similarity index as $\mathcal{J}(c_i, c_j) = \frac{\mathbb{L}(c_i \cap c_j)}{\mathbb{L}(c_i \cup c_j)}$.
Fig.~\ref{fig:coenroll_cdfV4}a shows Jaccard similarity index of
the top-$N$ courses, with $N$ varying from 10 to 500.
The figure reveals that  99\% of course pairs among all pairs
are with a Jaccard similarity index less than 0.1.
In other words, there is a very small fraction of courses that are frequently
co-enrolled by learners.
To further identify chances that two courses are co-enrolled by learners, we quantify the co-enrollment of courses by point-wise mutual information to mitigate the influence of random course enrollments. 
Here, random course enrollment refers to learners enroll in courses just for casual experience or curiosity.

\subsection{Course Co-enrollment Patterns}
Point-wise mutual information (PMI) between a pair of courses $c_i$ and $c_j$ quantifies the discrepancy between the probability of their coincidence given their joint and individual distributions:
\begin{equation}\label{equ:pmi}
	pmi(c_i, c_j) = log(\frac{p(c_i,c_j)}{p(c_i)p(c_j)}),
\end{equation}
where $p(c_i)=\mathbb{L}(c_i)/\mathbb{L}(c)$, and $\mathbb{L}(c_i)$ is the number of learners enrolled in course $c_i$ and $\mathbb{L}(c)$ is the number of all learners that enroll in at least one course in the platform.
$p(c_i,c_j)=\mathbb{L}(c_i\cap c_j)/\mathbb{L}(c)$ is the probability that courses $c_i$ and $c_j$ are co-enrolled by learners.
To normalize the PMI value into a limited interval, we further use 
normalized point-wise mutual information:
\begin{equation}\label{equ:npmi}
	npmi(c_i, c_j) = \frac{pmi}{-\log p(c_i,c_j)} = \frac{\log[ p(c_i) p(c_j)]}{\log p(c_i,c_j)} - 1.
\end{equation}
If co-enrollment of $c_i$ and $c_j$ is at random, then we have $\log p(c_i,c_j)= \log[p(c_i) p(c_j)] \implies nmpi(c_i, c_j)=0$.
We manually set $npmi(c_i, c_j)\geq 0.6$ to filter out the randomly co-enrolled course pairs and draw the course pairs that co-enrolled with high probability in network topology, as shown in Fig.~\ref{fig:coenroll_cdfV4}b.

First, the figure shows courses within the same category have a higher probability to be co-enrolled by learners.
We examine the top 10 co-enrollment pairs and find all pairs are from the same category or offered by the same university/organization.
For example, ``LabVIEW basic courses'' and ``LabVIEW advanced courses'' are two computer courses offered by National Instruments.
Second, the courses with a large number of learners are usually co-enrolled in courses from multiple categories, which reflects the diversity in learners' knowledge space.
Third, we observe the courses within the same category may separate into a few sub-clusters.

\begin{figure}
	\centering
	\includegraphics[trim=10 10 110 10, clip, width=0.92\linewidth]{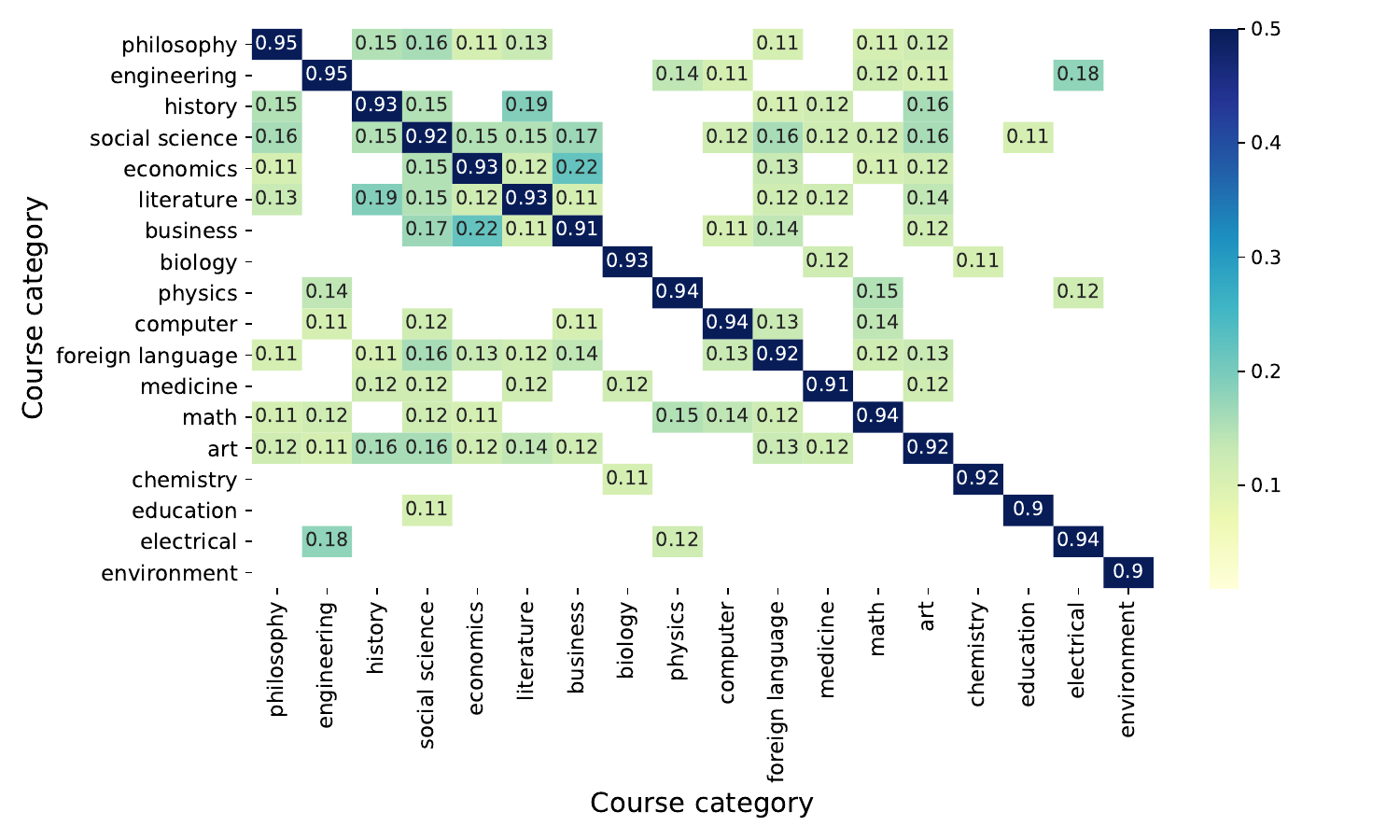}
	\caption{Heatmap of the category-level relationship of co-enrolled courses.}
	\label{fig:co_cat_htmp}%
    \vspace{-15pt}
\end{figure}

Our prior analysis reveals that a large proportion of course co-enrollment pairs are within the same category, we further study the correlation of course categories by computing the Jaccard similarity index between two course categories:
\begin{equation}\label{equ:cj}
	\mathcal{J}(C_i, C_j) = \frac{\mathbb{L}(C_i \cap C_j)}{\mathbb{L}(C_i \cup C_j)},
\end{equation}
where $\mathbb{L} (C_i)$ is the number of learners with at least one course enrollment.
In a special case where $C_i$ and $C_j$ are the same category \ie $C_i=C_j$, we
calculate:
\begin{equation}\label{equ:cj_same}
\mathcal{J}(C_i, C_i) = \frac{\mathbb{L}_2 (C_i)}{\mathbb{L} (C_i)},
\end{equation}
where $\mathbb{L}_2 (C_i)$ is the number of learners who enroll in at least two unique
courses from category $C_i$.

Fig.~\ref{fig:co_cat_htmp} shows the Jaccard similarity index between
each course category. 
It is surprising that the probability for learners enrolled in at least two courses in the same category is more than 90\%, as shown in the diagonal of the heatmap.
\xr{This finding appears counterintuitive, as one might expect a significant portion of learners to enroll in only one course within a specific category. However, considering Eq.\eqref{equ:cj_same}, this implies that over 90\% of learners enroll in at least two distinct courses belonging to the same category.}
The reason is that not all courses are associated with a category. For our category-level analysis, we only focus on courses tagged with a category. Most of these are popular courses among learners in a two-year period.
We also observe learners have a high probability to co-enroll in courses with related categories.
For example, the correlation between ``electrical'' and ``engineering'' is 0.18,
``business'' and ``economics'' is 0.22.
Additionally, the figure illustrates the diversity of learners' course enrollment.
For example, the learners majoring in the courses of ``business'' may also be interested in the courses of ``social science'' and ``computer'', while the crowd enrolled in the courses of ``social science'' may be also interested in ``art'' and ``philosophy''.

\subsection{Sequential Enrollment Pattern Mining}
\label{sec:seq_enroll}
From our temporal analysis, we notice the enrollments of courses for learners
are in a temporal order. This inspires us to view the course enrollment of the learners
as a sequence and study the course transition patterns.

\begin{figure}
	\centering
	\includegraphics[trim=10 10 110 10, clip, width=0.92\linewidth]{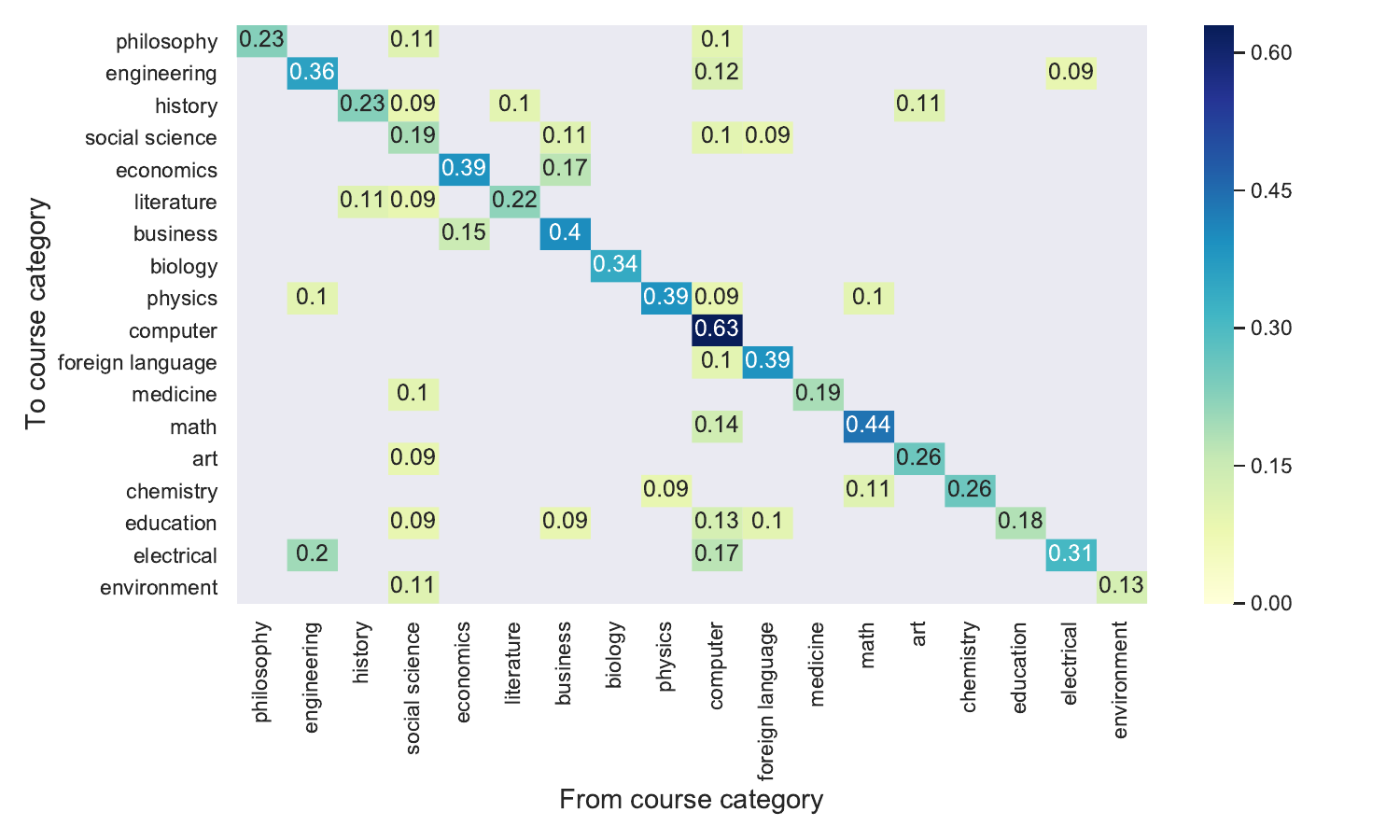}
    \vspace{-15pt}
	\caption{Category-level enrollment transitions.}
	\label{fig:cat_enroll_trans_htmp}%
    \vspace{-15pt}
\end{figure}

Analogous to Fig.~\ref{fig:co_cat_htmp}, we also plot the course transition at  category-level in Fig.~\ref{fig:cat_enroll_trans_htmp}.
Each row in the heatmap denotes the empirical distribution from one course category to all other course categories.
Given a set of transitions, $D=\{x_1, ..., x_k, ..., x_N\}, 1 \le k \le N$, where $N$ is the total number of transitions.
We denote a course enrollment transition $x_k$ that starts from course $m$ to course $n$ as $x_{k,m \to n}$.
The empirical distribution on row $i$ starting
with category $C_i$ is defined as follows:
\begin{equation}
	p_{emp}(C_i \to C_j) = \frac{1}{M} \sum^{N}_{k=1} \delta_{x_{k,m \to n}}(C_i \to C_j),
	\label{equ:trans_cat}
\end{equation}
where $M=\sum^{N}_{k=1} \delta_{x_{k,m \to *}}(C_i \to *)$ is the number of transitions initialized from category $C_i$.
$\delta_{x_{k,m \to n}}(C_i \to C_j)$ is the Dirac measure, defined by:
\begin{equation}
	\delta_{x_{k,m \to n}}(C_i \to C_j) = \left\{
	\begin{array}{ll}
		1 & m \in C_i \land n \in C_j; \\
		0 & m \notin C_i \lor n \notin C_j.
	\end{array}
	\right.
\end{equation}

The figure shows that transitions within the same category dominate all other transitions.
We find the category-level enrollment course transition is consistent with the category-level course co-enrollment plot in Fig.~\ref{fig:co_cat_htmp}.
However, the transition probability is asymmetric in Fig.~\ref{fig:cat_enroll_trans_htmp}.
We believe this finding would benefit researchers on sequential recommender systems.
\xr{
For instance, the category transition probability matrix depicted in Fig.~\ref{fig:cat_enroll_trans_htmp} serves as an illustrative example of prior knowledge that can be incorporated into intent-aware recommendation models (\eg \citep{li2021intention}). This integration could facilitate the model's ability to learn more accurate user intent transitions.
}
Finally, the heatmap shows that courses in hard sciences in
general hold a higher transition cohesion (\ie transition occurs within a category) than
the courses in soft sciences.
For example, the transition cohesion for ``computer'' and ``math'' are
much higher than that of ``education'' and ``environment''.

\section{Course Recommendation (RQ3)}
\label{sec:recom}
In this section, we exemplify how the findings of Section~\ref{sec:enrollment} can be
applied in manifold applications by addressing the last research question:
\textit{Can the observed patterns be used for course recommendation?}

\subsection{\xr{The Proposed Model: \textbf{FrePaPop}}}
\xr{\textbf{Motivation: }Building upon our prior findings, this work proposes a novel recommender system model that integrates sequential enrollment patterns with course popularity. This approach aims to leverage the insights gained from the observed course transition sequences, while also considering the broader appeal of courses across the platform. \\
\textbf{Model Design: }
This paper proposes FrePaPop, a recommender system model that leverages frequent course transition mining. In the first stage, FrePaPop identifies the most frequent subsequent course enrollments for each course within the training dataset. Analogous to Equation~\eqref{equ:trans_cat}, the model stores the transition probabilities for courses following each available course $c$. During the recommendation phase, given a learner's enrollment history culminating in course $c$, FrePaPop recommends the top-$K$ courses with the highest transition probabilities from $c$. If the number of available courses $(n)$ for recommendation falls below the threshold $K$ (\ie $n < K$), FrePaPop supplements the recommendation list with the top-$(K-n)$ most popular courses across the entire platform.}
\subsection{Datasets and Baselines}
\xr{
Our dataset encompasses learning activities from a two-year period. As evident from Fig.~\ref{fig:logs_daily}, the learning activity patterns exhibit a high degree of similarity between these two academic years. To facilitate computational efficiency and analysis, we extract data from learners with complete profiles for the academic year 2016-2017 for subsequent recommendation evaluation.  Within this year, the learning activities recorded from the first semester (August 1, 2016 - January 21, 2017) constitute the training set, while the data from the subsequent semester (January 22, 2017 - July 31, 2017) serves as the testing set.
The resulted dataset contains 20,752 learners and 1,104 courses, with
a sparsity level of 99.52\%.}
Several state-of-the-art methods are used as baselines for comparison.
1). \textbf{Random} algorithm recommends items randomly. 
2). \textbf{Popularity} recommends the most popular items.
3). \textbf{BiNE}~\citep{gao2018bine} is a bipartite network embedding-based method that recommends items based on the learner and course embedding. The edge weight denotes whether a learner enrolls in a course. 
4). \textbf{BiNE-Weight} initializes the edge weight in BiNE as the number of learning activities observed between a learner and his/her enrolled courses.
5). \textbf{FPMC}~\citep{rendle2010factorizing} recommends items by combining both sequential behavior and the general taste of users.
6). \textbf{Course2vec} is an algorithm derived from word2vec~\citep{mikolov2013distributed}. Here, each learner and her enrolled courses are analogous to a sentence and words in the sentence. With the obtained embedding of courses, we calculate the embedding of a learner by averaging the embeddings of her enrolled courses. Finally, we recommend top courses to each learner by computing the Euclidean distance between the learner and all candidate courses.

\subsection{Experimental Results}
\xr{We present the performance of each evaluated method on all considered metrics, along with their training time, in Table~\ref{tab:evalu_recom}. }
The table shows that the sequential recommendation models (\ie FPMC and FrePaPop) outperform other baseline algorithms significantly.
Course2vec also achieves competitive performance to the graph embedding based method, BiNE.
One of the major advantages of course2vec, FrePapop and FPMC is their ability to capture the temporal course enrollment of the learners.
It is paradoxical that the performance of BiNE-Weight with more information embedded is worse than BiNE.
The edge weight information weakens the enrollment patterns of learners.
From the experimental results, we find that recommending the most popular courses to learners is a plausible choice in course recommendation
\xr{The time complexity of the popularity algorithm is an order of magnitude lower compared to course2vec and FrePaPop.}
BiNE is the most time-consuming model, because it is a general approach that not only models the local proximity between learners and courses but also captures the high-order proximity (implicit relations) between learners or courses.
It is worth noting our proposed FrePapop obtains competitive recommendation accuracy to FPMC with a reduction of more than 200$\times$ on training time.
\xr{During inference, FPMC and our proposed model require 5.8$s$ and 0.3$s$, respectively, to generate course recommendations. Consequently, FrePapop demonstrates significantly faster inference speed, making it more suitable for real-time recommendation scenarios.}

\begin{table}
	\small	
        \renewcommand{\arraystretch}{1.0}
	\caption[Table caption text]{Comparison of recommendation performance between our proposed FrePaPop and the baselines. We mark the global best results under each metric in \textbf{boldface} and the second best is \underline{underlined}.}	
	\label{tab:evalu_recom}
	\begin{tabular}{ l llll r}
		\toprule
		Methods & F1@10 & MAP@10 & MRR@10 & NDCG@10 & Training Time ($s$) \\
		\midrule
		Random & 0.63\% & 0.29\% & 1.28\% & 0.67\% & \textbf{0.1} \\
		Popularity & 3.96\% & 1.39\% & 4.38\% & 3.69\% & \underline{0.2} \\
		BiNE-Weight & 1.47\% & 1.07\% & 3.78\% & 1.17\% & 3881  \\
		BiNE & 3.95\% & 1.38\% & 4.31\% & 3.67\% & 4882 \\
		course2vec & 3.43\% & 1.94\% & 5.36\% & 3.87\% & 1.5 \\
		FPMC & \textbf{5.31\%} 
		& \textbf{3.72\%} & \textbf{10.57\%} & \textbf{6.81\%} & 221 \\
		FrePaPop (Our) & \underline{4.76\%} & \underline{3.66\%} & \underline{9.73\%} & \underline{6.42\%}  & 1.5  \\	
		\bottomrule
	\end{tabular}
    \vspace{-20pt}
\end{table}

%
\section{Implications and Limitations}
In this section, we discuss how the observed learning patterns can benefit relevant stakeholders including platform operators, course instructors, education researchers, and end-learners.
Then, we state the limitations
and threats to validate and generalize the results
in our study.
\subsection{Implications and Discussion}
\noindent\textbf{Improving course recommendation}.
Our prior experimental results reveal that
the sequential course enrollment pattern of the crowd can be
used as a practical guide for recommendation systems.
In addition, the category-level course transition in Fig.~\ref{fig:cat_enroll_trans_htmp}
can also be fused in recommendation systems.
Both ``FrePaPop'' and ``Popularity'' algorithms recommend the most popular
courses in the training set without differentiating the course category.
We argue that recommending algorithms can be boosted by considering
the courses that relevant to the learner's subject.
Due to the remarkable sparsity of the enrollment data, the performance of traditional recommendation models is usually lower than expected. Under this situation, it is worth considering the enrollment patterns.
\xr{For example, the incorporation of popularity trends,
as captured by models like PARE~\citep{PARE}, could potentially enhance sequential recommendation performance by leveraging the transition distributions identified in RQ2.
Moreover, the co-enrollment patterns identified in this work, particularly the correlations observed between course categories, offer promising avenues for improving the performance of various recommendation models~\citep{BM3,FREEDOM,DRAGON,DGVAE,MP4SR,DWSRec,WhitenRec}.}

\noindent\textbf{Untangling course prerequisite relations}.
Current research on course prerequisite relations mainly depended on course prerequisite data, labeled concept prerequisites or video playlist~\citep{pan2017prerequisite, lu2019concept}.
\xr{
While current prerequisite structures are meticulously curated by domain experts over extended periods~\citep{roy2019inferring}, they may not comprehensively reflect the dynamic nature of MOOC platforms that continuously evolve with the addition of thousands of courses.  To address this limitation, recent research suggests that course enrollment and transition patterns can be  exploited to automatically infer prerequisite relationships between courses or learning modules~\citep{pan2017prerequisite}.
}
\xr{
To illustrate, our analysis of co-enrollment patterns reveals frequent co-enrollment of course pairs, such as ``LabVIEW basic courses'' and ``LabVIEW advanced courses''. As shown in Fig.~\ref{fig:cat_enroll_trans_htmp}, category-level enrollment transitions between courses exhibit an asymmetry pattern. By identifying both the course opening timeline and course transition probability, we can readily identify prerequisite courses.
Consequently, course recommendation models (\eg~\citep{LayerGCN,SelfCF}) or learning path planning models (\eg~\citep{siren2022automatic}) can benefit from incorporating this prior knowledge of prerequisites into their models.
It is worth mentioning that learning activity data from a large number of users is readily obtainable, offering several orders of magnitude more data compared to labeled data.}

\subsection{Limitations}

\noindent\textbf{Single MOOC platform}. Our dataset was originally collected from
a single MOOC platform, XuetangX. Such a limitation may have
introduced some biases caused by the specific policies
in the platform \eg the sequential enrollment pattern
observed in the subsection~\ref{sec:seq_enroll}.
In XuetangX, learners can enroll an already ended course,
thus, the patterns in other platforms may be different from our observations.
With the different available courses hosted on the platform, the co-enrollment
pattern may also slightly differ from our findings.
However, the methodologies used in our analysis can be applied to datasets from other platforms.

\noindent\textbf{Learners' demographics}.
Learners engaged in this platform are mainly from mainland China, hence, regional factor could affect the results in the analysis.
It is reported that learners from different countries or associated cultures exhibit different learning behaviors in MOOCs~\citep{liu2016mooc}.
Besides, the conventional course schedule in brick-and-mortar colleges from different countries may also influence the temporal patterns of learners.
Previous researches~\citep{qiu2016modeling,bayeck2016exploratory} found that learners' retention and completion of the course are correlated with their demographics.
However, our study on millions of learners reveals some common learning patterns that can be shared across different platforms or countries.

\section{\xr{Literature Review}}
\xr{In recent years, the emergence of MOOCs leads to the availability of large-scale educational datasets collected from a diverse range of learners.
Various studies have examined different aspects of educational data.

\textbf{Demographic factors}. \cite{qiu2016modeling} conducted a regression analysis on 11 courses to examine the correlation between the learners' learning patterns and their demographics.
 They demonstrated that female and male learners exhibited significantly different learning patterns. Female learners were more likely to engage in non-science courses. Learners with a bachelor's degree asked more questions, while graduate learners were more active in answering questions.
Other studies~\citep{guo2014demographic, liu2016mooc, bayeck2016exploratory, aljohani2021learners, ruiperez2022large} also found that demographic differences can greatly influence the learning patterns of MOOC participants. For example, learners from East Asia tended to take more assignments, while learners from Australia, Canada, the U.S., and the U.K. showed moderate engagement in video watching and assignment solving.
In addition to these relationships, researchers have aimed to predict MOOC learner engagement and completion rates. \cite{morris2015can} analyzed the key demographic factors that influence course completion. They found a significant association between learners' age, prior MOOC experience, highest level of education, and employment status with the degree of completion.
However, the authors of \citep{whitehill2017mooc} examined 40 online courses from HarvardX and concluded that demographics provided only a limited amount of information about dropout. The study by \citep{feng2019understanding} revealed that female learners had a slightly higher dropout rate in hard science courses (\eg math, physics, etc.). The authors argued that learning patterns played an important role in predicting completion rates, which is consistent with the findings of~\citep{coffrin2014visualizing, brooks2015you}. The authors in \citep{feng2019understanding} also suggested that learners' social networks in forums were a significant feature for dropout prediction.

\textbf{Learning behavior analysis}.
\cite{jiang2015learning} analyzed six courses on the Coursera platform and found that the learning behavior of learners is associated with the completion of a course.
The temporal pattern shown in Chinese courses differs from English courses.
From the learning behavior analysis, they revealed the certificate achieved learners are those who watched the video multiple times and highly active in assignments.
A few studies~\citep{liang2014analysis, wang2015investigating, kennedy2015predicting, coffrin2014visualizing, henderikx2017refining, wise2018unpacking} also examined how learners' learning behavior would affect MOOC learning performance.
Building on Maslow's hierarchy of needs theory, the MPBN model~\citep{zhu2023mooc} leveraged the OULAD dataset~\citep{kuzilek2017open} to predict learner performance.
Research~\citep{xie2024did} also delved into the intrinsic and extrinsic motivations behind repetitive  learners engaging with MOOC platforms.
\cite{zhang2017smart} analyzed how learners navigating through a course video and proposed a predicting model to support learners' learning experience in watching videos.
The researches of~\citep{matcha2019analytics} and~\citep{xu2023participation} are similar to our research in the analysis of learning trace data.
Clustering and sequence mining methods are applied to detect the learning tactics of learners. However, their study only focus on single course.

In summary, our study differs from previous work significantly in terms of scale, and research perspectives (Refer Table~\ref{tab:dtcomp} for a comprehensive comparison of the available MOOC datasets).}

\begin{table}
	\footnotesize
	\setlength\tabcolsep{1.2pt} 
        \def\arraystretch{0.9}	
	\caption{\xr{Comparison of current available MOOC datasets on scale and research topics.}}
	\label{tab:dtcomp}
	\begin{tabular}{ l | l | l | c | l }
		\hline
		Literature & Courses & Learners & Learning logs & Research Perspectives  \\
		\hline
		\cite{guo2014demographic} & 4 & 141K & \cmark & Learning patterns $w.r.t$ demographics  \\
		\cite{coffrin2014visualizing} & 2 & 92K & \cmark & Visualizing learning patterns  \\
		\cite{liu2016mooc} & 1 & 29K & \cmark & Learning patterns $w.r.t$ demographics \\
		\cite{bayeck2016exploratory} & 1 & 150K & \xmark & Course enrollment motivation $w.r.t$ demographics \\
		\cite{whitehill2017mooc} & 40 & 528K & \cmark & Mooc dropout prediction \\
		\cite{jiang2015learning} & 6 & 80K & \cmark & Learning behaviors $w.r.t$ performance  \\	
		\cite{kennedy2015predicting} & 1 & 37K & \cmark & MOOC performance $w.r.t$ prior knowledge etc.\\	
		\cite{henderikx2017refining} & 2 & 8K & \xmark &  MOOC performance $w.r.t$ learning behaviors \\	
		\cite{matcha2019analytics} & 1 & 1K & \cmark  &  Learning performance $w.r.t$ learning strategies\\	
		\cite{qiu2016modeling} & 11 & 88K & \cmark & Learning performance $w.r.t$ learning behaviors \\	
        \cite{kuzilek2017open} & 22 & 33K & \cmark &  Introducing of dataset\\	
            \cite{ruiperez2022large} & 6,111 & 8.67 M & \xmark & Differences $w.r.t$ demographics across MOOC providers \\
		\cite{xu2023participation} & 1 & 425 & \cmark & Learner participation types $w.r.t$ connectivism \\	
		\hline	
		\textbf{Out study} & 1,629 & 773K & \cmark & Temporal \& Co-enrollment Learning Patterns \\	
		\hline
	\end{tabular}
 \vspace{0pt}
\end{table}

\section{Conclusion}
This paper studies learners' learning patterns in a large-scale platform that hosts 351 million learning activities from 0.7 million learners in two years. The study is centered around learners' temporal and enrollment patterns to uncover the characteristic of the crowd in MOOCs.
We first show that learning activities of the crowd show saliently periodic patterns. The \xr{daily} and weekly patterns reflect the temporal stability and consistency in human behaviors. On top of a large number of courses, we analyze the course enrollment patterns of the crowd and find mostly co-enrolled course pairs belong to the same category or the same university. Our sequential pattern mining on course enrollment transition uncovers that hard sciences courses have a higher probability for the inter-category transition.
In the end, we demonstrate how our findings can boost manifold application, using a recommendation task as a case study.


\bibliographystyle{apacite}
\bibliography{interactapasample}

\end{document}